# Landau-Zener Transition in Topological Acoustic Pumping

Ze-Guo Chen[1], Weiyuan Tang[1], Ruo-Yang Zhang[2], and Guancong Ma[1,*]

[1]*Department of Physics, Hong Kong Baptist University, Kowloon Tong, Hong Kong*
[2]*Department of Physics, The Hong Kong University of Science and Technology,*
*Clear Water Bay, Hong Kong*

Quantized charge pumping is a robust adiabatic phenomenon uniquely existing in topologically nontrivial systems. Such topological pumping not only brings fundamental insights to the evolution of states under the protection of topology but also provides a vital approach for the study of higher-dimensional topological phases. In this work, we demonstrate topological pumping in an acoustic waveguide system by showing the successful pumping of topological boundary states (TBSs) across the bulk to the opposite edge. By exploring the finite-size induced mini-gap between two TBS bands, we unveil the quantitative condition for the breakdown of adiabaticity in the system by demonstrating the Landau-Zener transition with both theory and experiments. Our results not only open a new route for future investigations of topological acoustics but also establish non-adiabatic transitions as a new degree of freedom for manipulating topological states.

The emergence of topological effects in physics is often underlain by adiabatic cyclic variation of a certain set of parameters, which leads to quantized Berry phases that give rise to various types of topological invariants [1,2]. Perhaps the best-known example is a Chern insulator, which is characterized by a quantized topological invariant known as the TKNN number (or in many cases, the Chern number) [3]. Topological boundary states (TBSs) exist at the interface between two topologically distinct systems, which are immune to backscattering by defects and impurities, leading to robust unidirectional transport. Underpinned by the fundamental universality of topology, recent efforts have identified that similar physical phenomena not only exist in solid-state electronic systems, but also in photonics and electromagnetism [4,5], elastics and acoustics [6,7].

Quantized transport of charges is another important manifestation of topology [8]. A famous example is the "Thouless pumping," which describes the transport of a quantized amount of charges induced by an adiabatic cyclic variation of one-dimensional (1D) lattice potential [9]. It is regarded as a dynamical version of the two-dimensional (2D) quantum Hall effect [10-14]. Thouless pumping has been observed using optical lattice [15-17], photonic waveguide arrays [18-21], and elastic plates [14,22]. Meanwhile, topological pumping offers a unique platform for investigating the topology of higher-dimensional physics [19,23]. Although topological pumping is widely regarded as a robust phenomenon, the quantized pumping can only be well observed under the adiabatic condition that requires a sufficiently slow variation of phase-space parameters. Deviation from adiabatic approximation will inevitably populate other states and eventually break the quantized property of the final state. However, despite its fundamental importance, the occurrence of non-adiabatic transition in topological pumping so far remains largely unexplored.

In this letter, we present a theoretical and experimental study of non-adiabatic transition in dynamic topological pumping in acoustics. The acoustic topological pumping is based on the TBSs existing in a commensurate Harper model [24]. The two TBSs, localized at the two opposite boundaries of a finite chain, constitute a two-level system on which our dynamic pumping is enforced. Importantly, the two TBSs can form a small gap by finite-size induced avoided crossing [25], which we exploit for the investigation of the breakdown of adiabaticity. We reveal with theoretical and numerical analyses that such a breakdown is indeed induced by the Landau-Zener transition [26] in the TBS pumping, wherein the adiabatic condition is linked to the rate of parameter modulation as well as the size of the TBS gap. Our acoustic experiments not only successfully observed topological pumping in acoustics, which has yet been realized to date, but also confirmed the non-adiabatic transition described by the Laudau-Zener model. Our results offer the necessary insights to future applications relying on the adiabatic evolution of states and bring new possibilities for acoustic, elastic, and electromagnetic devices exploiting non-adiabatic transitions.

We begin by considering a tight-binding finite chain shown in Fig. 1(a), described by a commensurate Harper model [24]

$$H(\phi) = \sum_{m}^{N}[f_0 + \sigma\lambda\cos(2\pi bm + \phi)]|m\rangle\langle m| + \sum_{m}^{N-1} t|m\rangle\langle m+1| + \text{h.c.,} \quad (1)$$

where $m$ labels the sites, $t$ is the hopping coefficient, $N$ is the total number of sites in the finite chain, $\sigma = \pm 1$ is a degree of freedom. The onsite terms are periodically modulated as $f_0 + \sigma\lambda\cos(2\pi bm + \phi)$. Here, $\lambda$ is the modulation amplitude and $b = \frac{1}{3}$ is the modulation spatial frequency, and the parameter $\phi$ serves as a synthetic dimension of the system. Two eigenspectra are plotted in Fig. 1(b) as functions of $\phi$. Both spectra show three bulk band regions separated by two bandgaps, which are bridged by two chiral TBSs that are protected by nontrivial Chern number $C$ computed in $k\phi$-plane, as labeled in Fig. 1(b). Although two chiral TBSs are found in both cases, their relative locations on the $\phi$-axis are drastically different. When the total site number is $N = 3n$ with $n$ being an integer, the two TBSs meet, and a small gap is opened due to the avoided crossing induced by a finite-sized effect of the chain (left panel) [27]. But if the total site number is $N = 3n - 1$, the two TBSs are symmetric about $\phi = 0$ and they do not encounter (right panel).

The tight-binding chains can be realized using an acoustic waveguide array [28,29], as schematically shown in Fig. 1(c) and (d). The system consists of $N$ rectangular air-filled waveguides coupled by a thin sheet of air at the bottom. The cross section of the array reproduces the Harper model (Eq. (1)) with a specific $\phi$. Here, we employ the first-order guiding mode, which we write as $|\psi_j(\phi)\rangle e^{ik_z z}$, where $|\psi_j(\phi)\rangle$ is an eigenfunction of $H(\phi)$ with a corresponding eigenfrequency $f_{H,j}(\phi)$, with $j$ labeling the bands. When the working frequency is fixed at $f_w$, the propagating wave number is $k_{z,j}(\phi) = \frac{2\pi}{c}\sqrt{f_w^2 - f_{H,j}^2(\phi)}$, where $c$ is the speed of sound. Equation (1) can then be transformed to

$$\mathcal{H}(\phi) = \Psi K_z \Psi^\dagger. \tag{2}$$

Here, $\Psi = (|\psi_1(\phi)\rangle, |\psi_2(\phi)\rangle \dots |\psi_N(\phi)\rangle)$ is a column matrix formed by all eigenvectors of $H(\phi)$; $K_z = \sum_j^N k_{z,j}(\phi)|j\rangle\langle j|$. Equation (2) shares the same set of eigenfunctions as Eq. (1), but the corresponding eigenvalues become $k_{z,j}$, as shown in Fig. 1(e). It follows that $\phi$ directly links to the eigenspectrum $k_z$ of the waveguides. We can then enforce the adiabatic variation of $\phi$ by slowly modulating the waveguide along $z$-direction, which is described by

$$-i\partial_z|\psi(z)\rangle = \mathcal{H}(z)|\psi(z)\rangle, \tag{3}$$

which is mathematically equivalent to the Schrödinger equation but with the time replaced by $z$. Since the first-order guiding mode resembles a dipole in its cross-sectional profile, the adiabatic variation of $\phi$, which affects the onsite frequency (Eq. (1)), can be implemented by a continuous change of the height of each waveguide, as schematically shown in Fig. 2(a).

As shown in Fig. 1(b, f), the adiabatic variation of $\phi$ can drive the TBS localized at one boundary (state A) to the opposite boundary (state C), realizing a topological pumping. The acoustic waveguide array offers a way to achieve topological pumping by continuously changing the height of each waveguide along the propagating direction $z$. Although the end states in Fig. 1(b, f) are essentially the same, the difference between the two cases, respectively containing $3n$ and $3n - 1$ waveguides implies two distinct pumping processes. When the system contains $3n - 1$ waveguides, the transition between two TBSs must connect through a bulk band near $\phi = 0$ or $\pi$. This implies that $\phi$ must be driven across a rather large range in $\phi$ so that the initial and final states both locates away from the bulk band, which is necessary for them to be well localized before and after the pumping. In reality, it means the system must be sufficiently large, as we show in [30]. On the contrary, when the system contains $3n$ waveguides, the TBSs cross at $\phi = -\pi/3$ (in lower gap). When the number of sites is relatively small, the two crossing TBSs, despite localized at two opposite boundaries, couple evanescently and result in avoided crossing, leading to a small gap opened near the crossing points [27]. In this case, the TBS can be pumped to the opposite boundary directly through tunneling. Such tunneling can occur within a much smaller range of $\phi$ and without involving any bulk states, as shown in Fig. 1(f). However, as we will demonstrate next, the small size of the TBS gap means that such tunneling effect is sensitive to the condition of adiabaticity and therefore offers a unique opportunity for studying non-adiabatic transition in acoustic topological pumping.

To investigate the non-adiabatic transition, we use a waveguide array that contains 9 waveguides. The height of each individual waveguides is modulated according to $h_m = 20 + 5\cos(2m\pi/3 + \phi)$ mm. This is schematically shown in Fig. 2(a). The hopping among waveguides is achieved by a layer of air with a thickness $w = 3$ mm so that TBS gap is easily observable. To compensate for the perturbation to onsite frequency induced by the coupling, the two waveguides at the boundaries have an additional height correction of $\Delta h = -2$ mm [31]. Mode analysis using the finite-element software (COMSOL Multiphysics) at a working frequency $f_w = 9.5$ kHz reveals that the TBS gap has a size of $\Delta k_z = 4.66$ m$^{-1}$, as shown Fig. 2(b). In the vicinity of $\phi = -\pi/3$, the avoided crossing of the TBSs can be modeled by a two-level effective Hamiltonian near $k_z = 53.7$ m$^{-1}$

$$H_e(\delta\phi) = \begin{pmatrix} -\alpha\,\delta\phi & \Gamma \\ \Gamma & \alpha\,\delta\phi \end{pmatrix}. \tag{4}$$

Here, $\Gamma = \Delta k_z/2$ is determined by the gap size and $\alpha = 70.4$ m$^{-1}$ is a fitting parameter. The bases of $H_e$ are $|\psi_L\rangle$ and $|\psi_R\rangle$, i. e., the TBSs localized at the left and right boundaries, respectively. The eigenvalues of $H_e$ are plotted as the solid curve in Fig. 2(b), which are in excellent agreement with the results from mode analysis (dots). The weighting of $|\psi_L\rangle$ and $|\psi_R\rangle$ is shown by the blue and red colors. The initial state is $|\psi_R\rangle$ at $\phi = -0.38\pi$ (marked by red star). The two possible routes of state evolution across the TBS gap are represented by the dashed arrows in Fig. 2(b). The final state $|\psi_f\rangle$ is a combination of two boundary states $|\psi_R\rangle$ and $|\psi_L\rangle$. When the pumping follows the red path, $|\psi_R\rangle$ component dominates during the pumping process, and the final state remains localized on the right boundary. Alternatively, when it is pumped along the blue path, the $|\psi_L\rangle$ component dominates the final state which induces field localization on the left boundary. These are plotted in Fig. 2(c).

The composition of the final state can be predicted by the Landau-Zener model [26,32], which gives the condition of non-adiabatic transition in the pumping. The final state depends on the TBS gap size $\Delta k_z$ and the pumping rate. In our system, the gap size is apparently fixed. The pumping rate is determined by $\Delta\phi/z_m$, i.e., the ratio of the total pumping range $\Delta\phi = 0.1\pi$ and the length of the modulated waveguides $z_m$ (the red section in Fig. 2(a)). Then the composition of the final state is given by $|\psi_f\rangle = L(z)|\psi_L\rangle + R(z)|\psi_R\rangle$, with $L(z) = \langle\psi_f|\psi_L\rangle$ and $R(z) = \langle\psi_f|\psi_R\rangle$ satisfying

$$-i\frac{d}{dz}\begin{pmatrix}L(z)\\R(z)\end{pmatrix} = \begin{pmatrix}\beta z & \Gamma\\ \Gamma & -\beta z\end{pmatrix}\begin{pmatrix}L(z)\\R(z)\end{pmatrix}, \quad (5)$$

where $\beta = \alpha\frac{\Delta\phi}{z_m}$ characterizes the adiabaticity. In our system, the pumping begins with the state dominantly at $|\psi_R\rangle$ so that the initial condition is $(0, 1)^T$. From Eq. (5), we can work out the final state as a function of $z_m$, i. e., $R^2(z_m) = |\langle\psi_f|\psi_R\rangle|^2 = e^{-\pi\Gamma^2/\beta}$ and $L^2(z_m) = 1 - e^{-\pi\Gamma^2/\beta}$. We plot the final state weighting in Fig. 2(d) as functions of $z_m$. It is seen that $R^2(z_m)$ and $L^2(z_m)$ cross at

$$z_t = \frac{\alpha\,\Delta\phi\ln 2}{\pi\Gamma^2}. \quad (6)$$

We denote this to be the point of Landau-Zener transition. The physical indication of this transition can be understood straightforwardly. With the pumping range $\Delta\phi$ is fixed, the length of waveguides determines the pumping rate. When $z_m \gg z_t$, i.e., the waveguide array is very long so that the state evolution is sufficiently slow, the process is adiabatic, and the state remains on the same band throughout the pumping. This means that the pumping follows the blue path in Fig. 2(b). Consequently, the final state is dominated by the TBS localized at the left boundary. However, when the same pumping is enforced with a short waveguide array with $z < z_t$, the variation is fast enough so that adiabaticity is not satisfied. Consequently, the upper TBS band is excited, and the pumping follows the red path in Fig. 2(b). In this case, the final state has a large component of $|\psi_R\rangle$, i.e., the energy remains largely localized at the right boundary. The energy of the final state redistributes according to the pumping rate, which is also clearly verified through solving Eq. (3) dynamically, as shown in supplemental materials [30].

In our acoustic designs, the transition length is $z_t = 0.899$ m, which is experimentally feasible for our lab. We simulate four configurations to verify our analytical prediction. The numerical model is schematically shown in Fig. 3(a). The simulated result of configuration I ($z_m^I = 0.2$ m, Fig. 3(b)) shows that the initial and final states are both the right-residing TBS, whereas for configuration II ($z_m^{II} = 0.89$ m), the result shows a precisely equal distribution of acoustic energy at the left and right boundaries (Fig. 3(c)). For configuration III ($z_m^{III} = 1.6$ m), the majority of the acoustic energy in the final state is localized at the left boundary, as shown in Fig. 3(d). Further increase in $z_m$ leads to the domination of $|\psi_R\rangle$ appearing the final state, as seen in Fig. 3(e), where a longer modulation length $z_m^{IV} = 2.5$m is used. These results tally well with the theory. It is also noteworthy that in configurations II, III, and IV, the energy indeed is tunneled through the bulk to excite the left TBS without involving any bulk states, which is anticipated in our analysis.

Experimentally, the acoustic waveguide arrays were precision-machined from a block of aluminum. We fabricate two configurations of waveguide arrays with different modulation lengths: $z_m^I = 0.2$ m and $z_m^{III} = 1.6$ m, as shown in Fig. 4(a). We excite the initial state $|\psi_R\rangle$ at waveguide 9 at the input end, as shown in Fig. 4(b). To ensure that the first guiding mode is correctly excited, we use two anti-phase loudspeakers to drive a short section of unmodulated waveguides with a length of 0.1 m, which is connected to the modulated waveguides. The intensity profile at the output end is measured using a microphone. For configuration I with $z_m^I < z_t$, the output profile shows that the acoustic energy resides at the right boundary (Fig. 4(c)), indicating that the final state is dominantly $|\psi_R\rangle$. In contrast, for configuration III with $z_m^{III} > z_t$, a majority portion of acoustic energy is found transferred to the left edge, as seen in Fig. 4(d). The constraints in both fabrication capability and laboratory space prevent the experiments on even longer waveguides. Nevertheless, our results already demonstrate the successful acoustic topological pumping and the non-adiabatic transition that follows the Landau-Zener model.

In conclusion, we have not only demonstrated topological pumping in acoustics for the first time but also thoroughly investigated the non-adiabatic transition arising therein, which is described by a two-band Landau-Zener model. By exploiting the crossing of TBSs in the Harper model, our scheme can pump the TBS across the bulk to the opposite boundary by a small range of modulation. The entire process does not involve bulk states, making it more resilient against imperfections and loss [18]. Our approach of acoustic topological pumping can be extended to higher dimensions and can benefit other physical systems, such as mechanical vibrations, elastic waves, electromagnetism, electrical circuitries, and thermal transfers. The quantitative identification of the non-adiabatic transition condition offers insights into the adiabaticity in dynamic pumping in various systems. In particular, our work may lead to the reliable control of non-adiabatic transitions in non-Hermitian systems which is a major hurdle preventing the visualization of exotic topology by the dynamic encircling of exceptional points [33-35]. We also expect our work to open up more directions on topological waves with the new potentials for topological devices. The Landau-Zener model offers a new route for the precise control and manipulation of acoustic energy distribution in TBS pumping, with potential applications such as the Landau-Zener-Stückelberg interferometry [36], asymmetric guiding mode switching [35], wave splitting, multiplexing and de-multiplexing of waveguide channels [37], and so on.

**Acknowledgements.** This work was supported by Hong Kong Research Grants Council (GRF 12302420, 12300419, ECS 22302718, CRF C6013-18G), National Science Foundation of China Excellent Young Scientist Scheme (Hong Kong & Macau) (#11922416) and Youth Program (#11802256), and Hong Kong Baptist University (RC-SGT2/18-19/SCI/006). The authors wish to thank C. T. Chan, Zhao-Qing Zhang and Wei-Wei Zhu for helpful discussions.

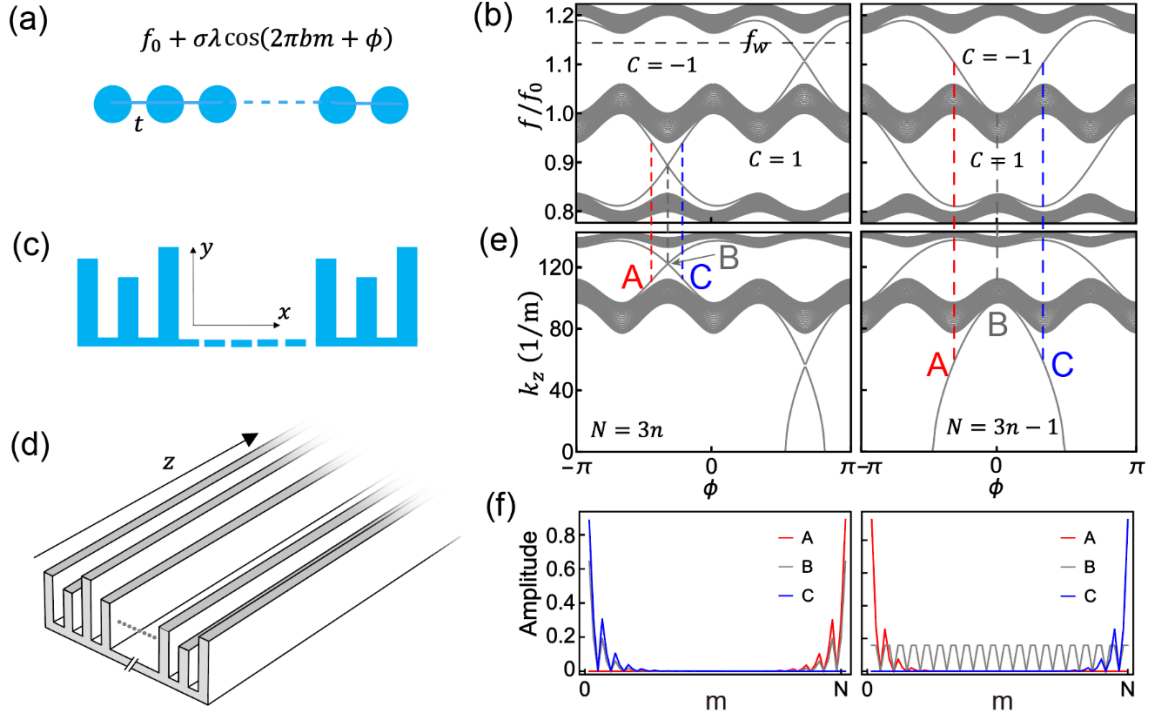

FIG. 1 (a) A schematic of a finite Harper chain. The onsite frequency is modulated according to the shown formula. (b) The calculated eigenspectra as functions of modulation phase ϕ based on Eq. (1). The parameters are: $f_0 = 9101.5$ Hz, $t = -0.082 f_0$, $\lambda = -2t$, $f_w = 1.16 f_0$. The left panel shows the result of a chain with $N = 60$ sites and $\sigma = -1$. For the right panel, $N = 59$, $\sigma = 1$. (c), (d) Schematics of an acoustic waveguide array for realizing the finite Harper chain. (e) The calculated eigenspectra of the Hamiltonian in Eq. (2) as functions of modulation phase ϕ, in which the eigenvalues are the propagating wave number $k_z$. The states at A, B, and C in (e) are shown in (f).

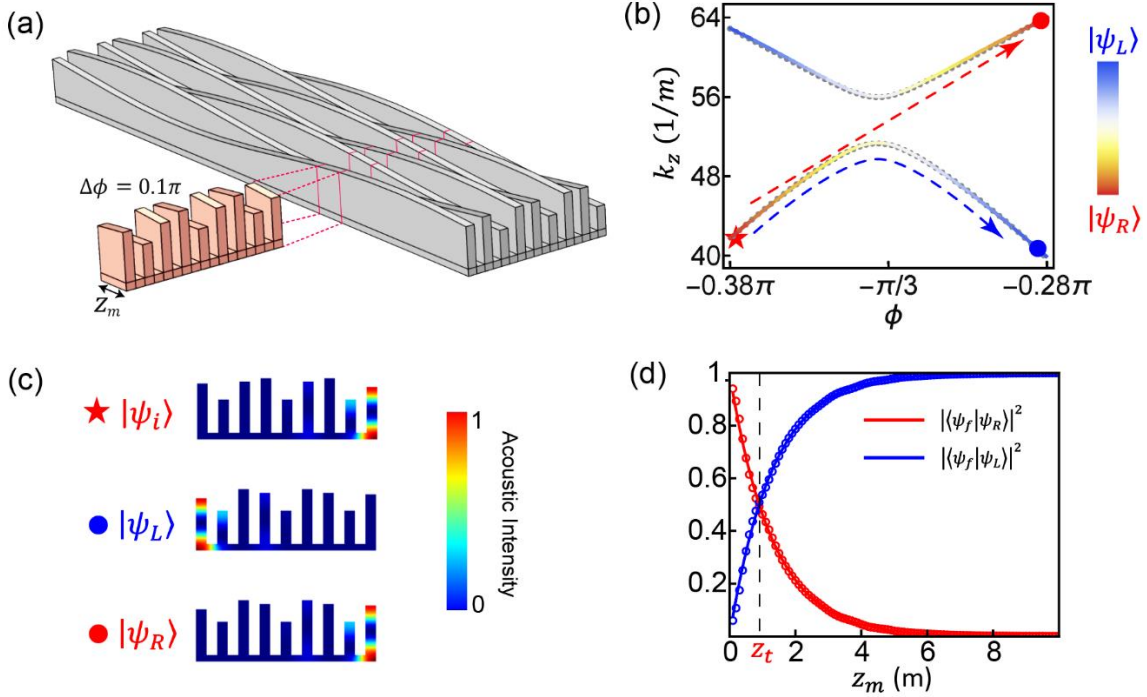

FIG. 2 (a) The acoustic waveguides array with modulated height along the $z$ direction. The modulation range in our investigation is $\Delta\phi = 0.1\pi$ near the TBSs crossing point $\phi = -\pi/3$, which is shown by the red section. The schematics here are not drawn according to the real scale. (b) Propagation wave number $k_z$ as a function of $\phi$ near $\phi = -\pi/3$ at the working frequency $f_w = 9.5$ kHz. The solid curves are results based on effective two-band model $H_e$. The dots are the results of finite-element simulations. (c) The field distribution of the initial state (star) and the two possible final states (circles). (d) The weighting of left and right TBSs, $|\psi_L\rangle$ and $|\psi_R\rangle$, in final state $|\psi_f\rangle$ as functions of modulation length $z_m$. The solid curves are the direct results of the Landau-Zener model, and the open circles are from numerical simulations.

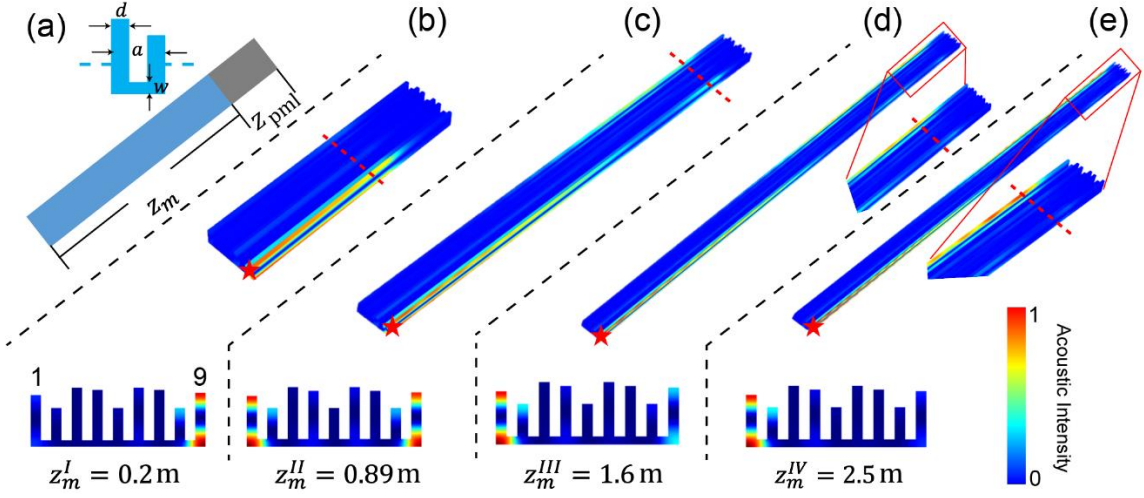

FIG. 3 (a) Schematics of the models used in finite-element simulations. Geometric parameters of the waveguide arrays on the *xy*-plane are: $d = 4$ mm, $w = 3$ mm and $a = 8$ mm. A length of perfect matched layer (PML) $z_{\text{pml}} = 0.1$ m is added to the output port to reduce the reflection. (b-d) Simulation results for configuration I (b), II (c), III (d), IV (e).

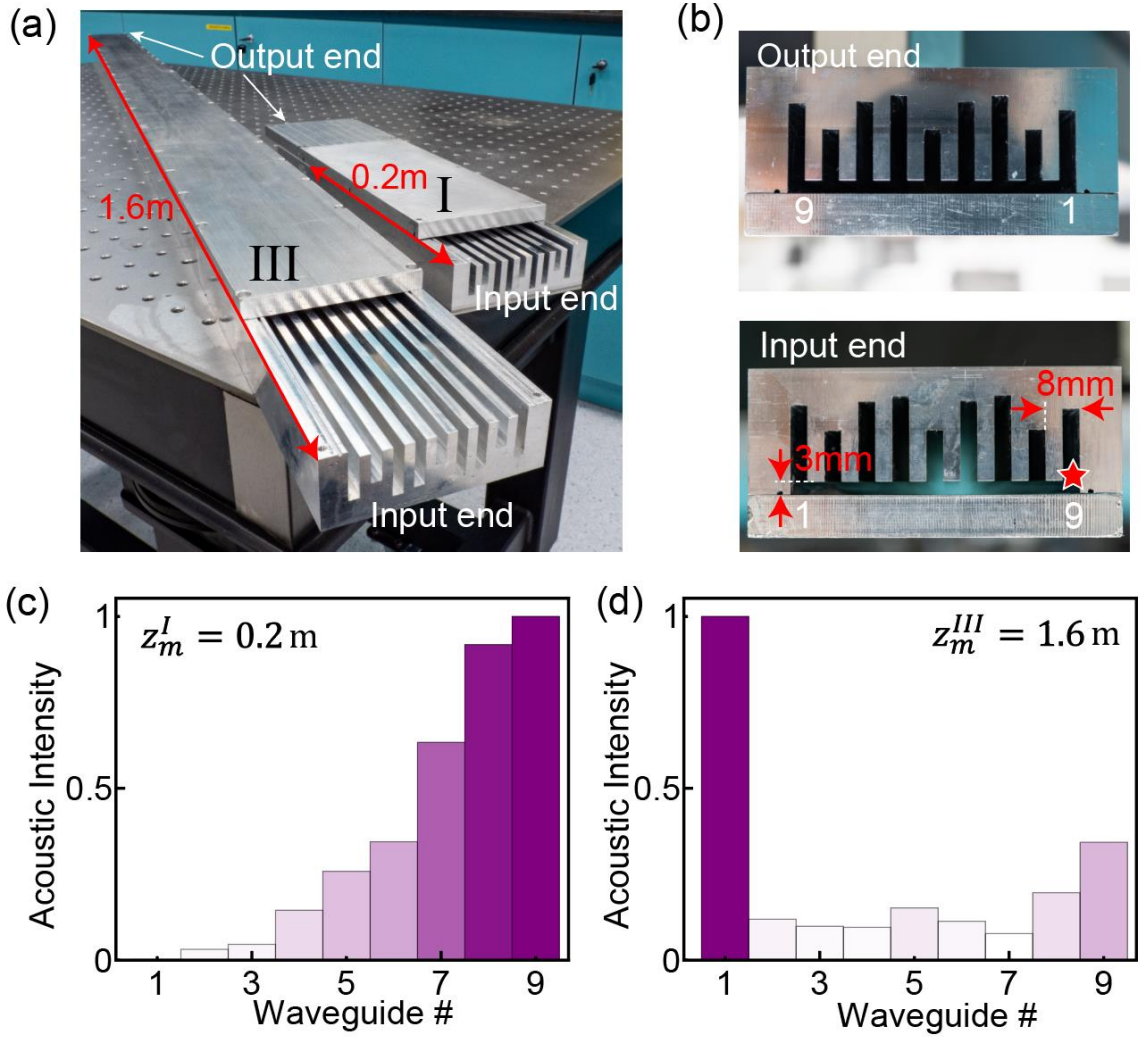

FIG. 4 (a) A photograph of the experimental acoustic waveguide arrays for configuration I and III. (b) Photographs of the input and output ends. (c) The measured acoustic intensity distribution at the output port for configuration I indicates that $|\psi_R\rangle$ dominates the final state. (d) The results for configuration III show a majority of acoustic energy is tunneled to $|\psi_L\rangle$ by the pumping. The results are in good agreement with our predictions.